\documentclass[12pt]{article}
\usepackage{amsmath,color}
\usepackage{amsthm}
\usepackage{graphicx,psfrag,epsf}
\usepackage{enumerate}

\usepackage{natbib}
\setcitestyle{aysep={}}
\usepackage{url} 
\usepackage{amssymb}
\usepackage{multirow}
\usepackage{array}
\usepackage{bm}
\usepackage{booktabs}
\usepackage{float}
\usepackage[font=small,labelfont=bf]{caption}

\usepackage[total={6.4in,8.6in}]{geometry}
\newcommand{\blind}{0}

\usepackage{setspace}
\newtheorem{theorem}{Theorem}

\newtheoremstyle{exampstyle}
{\topsep} 
{\topsep} 
{} 
{} 
{\bfseries} 
{.} 
{.5em} 
{} 

\theoremstyle{exampstyle}

\def\bSig\mathbf{\Sigma}

\usepackage{mathtools}
\usepackage{amsfonts}
\usepackage{multirow}

\DeclarePairedDelimiter\norm{\lVert}{\rVert}

\newcommand{\Real}{\mathbb{R}}
\newcommand{\V}[1]{{\bm{\mathbf{\MakeLowercase{#1}}}}} 
\newcommand{\M}[1]{{\bm{\mathbf{\MakeUppercase{#1}}}}} 
\newcommand{\argmin}{\operatornamewithlimits{arg\ min}}
\newcommand{\argmax}{\operatornamewithlimits{arg\ max}}

\begin{document}
	
	\def\spacingset#1{\renewcommand{\baselinestretch}%
		{#1}\small\normalsize} \spacingset{1}
	%
	%
	\if0\blind
	{
		\title{\bf Adaptive Semi-Supervised Inference for Optimal Treatment Decisions with Electronic Medical Record Data}
		\author{Kevin Gunn, Wenbin Lu and Rui Song
			\\
			Department of Statistics, North Carolina State University 
		}
		
		\date{\empty}
		\maketitle
	} \fi

	\baselineskip=21pt

\begin{abstract}
\indent A treatment regime is a rule that assigns a treatment to patients based on their covariate information. Recently, estimation of the optimal treatment regime that yields the greatest overall expected clinical outcome of interest has attracted a lot of attention. In this work, we consider estimation of the optimal treatment regime with electronic medical record data under a semi-supervised setting. Here, data consist of two parts: a set of `labeled' patients for whom we have the covariate, treatment and outcome information, and a much larger set of `unlabeled' patients for whom we only have the covariate information. We proposes an imputation-based semi-supervised method,  utilizing `unlabeled' individuals to obtain a more efficient estimator of the optimal treatment regime. The asymptotic properties of the proposed estimators and their associated inference procedure are provided. Simulation studies are conducted to assess the empirical performance of the proposed method and to compare with a fully supervised method using only the labeled data. An application to an electronic medical record data set on the treatment of hypotensive episodes during intensive care unit (ICU) stays is also given for further illustration. 
\end{abstract}

%


\newpage
\spacingset{1.45} 


%

\section{Introduction}
\label{intro}

\indent Precision medicine, which is focused on developing individualized  treatment decision rules by utilizing patients' covariate information, has earned considerable interest. A treatment decision rule is a deterministic function that assigns a treatment to patients according to their covariates. One main objective of precision medicine is to find the optimal treatment rule (OTR) that maximizes patients' expected outcomes of interest. There is a great deal of work on estimation of optimal treatment rules from randomized clinical trials or observational studies  \citep[e.g.][]{watkins1992q,murphy2003optimal,robins2004optimal,moodie2007demystifying, zhao2012estimating, zhang2012robust,Fanetal2017}. 
\\
\indent A recent area of interest in estimation of optimal treatment rules has been the utilization of Electronic Medical Record (EMR) data. For example, individualized treatment decisions are particularly suited to the ICU due to patients' heterogeneity for treating diseases like sepsis and acute kidney injury \citep{maslove2017path}. EMR data allow researchers to explore optimal treatment rules in specific clinical scenarios not feasible in randomized clinical trials (RCT) such as patients suffering from sepsis \citep{raghu2017continuous}. They can also provide guidance where there is little previous research conducted, such as second line treatment choices for type 2 diabetes episodes \citep{wang2016learning}. The recent shift in hospitals and other healthcare organizations to store clinical information in EMRs provide investigators with an immense amount of detailed information on the health status and clinical process of patients during interactions with their healthcare system. For example, MIMIC-III \citep{johnson2016mimic} is an openly available critical care database that encompasses medical record chart data recorded by clinicians, such as demographics, medical diagnoses, medications, vital signs, laboratory measurements (obtained in and out of the hospital), progress notes, imaging results, and billing information. The wealth of detailed clinical data provides the opportunity to enhance clinical decisions for accurate personalized medicine. 
\\
\indent Although secondary analysis of EMR data can serve as an important tool in improving patient outcomes, a major challenge in utilizing EMR data for estimation of OTR arises from the missingness and measurement errors in patients' records since EMR data are usually noisy and not recorded for research purposes \citep{weiskopf2013defining, beaulieu2018characterizing}. In this paper, we consider the estimation of OTR under a semi-supervised setting using EMR data. Specifically, data consist of two parts: a set of `labeled' patients for whom we have the covariate, treatment and outcome information, and a much larger set of `unlabeled' patients for whom we only have the covariate information. 

In classical semi-supervised learning, the knowledge gained through $\mathbb{P}_{\M{X}}$ (distribution of covariates) from `unlabeled' individuals is incorporated to improve inference on $\mathbb{P}_{\M{Y} \mid \M{X}}$ \citep{chapelle2009semi, zhu2009introduction}. When the amount of missing information far exceeds the amount of available information, semi-supervised learning techniques have been used for improving the prediction with EMR data. For example, \cite{chakrabortty2018efficient} proposed semi-supervised learning techniques to improve estimation efficiency of regression parameters in a linear working model using EMR data for studying the association between biomarkers and rheumatoid arthritis. \cite{beaulieu2016semi}, \cite{garla2013semi} and \cite{hong2018semi} developed semi-supervised methods for more accurate phenotype prediction when only a small subset of the patients underwent a manual chart review by a physician.

When estimating the OTR under the semi-supervised setting, we propose to use the covariate information in unlabeled data to derive the estimation of the contrast function, i.e.  $E\{\M{Y} | \M{X}, \M{A}=1\} - E\{\M{Y} | \M{X}, \M{A}=0\}$, which in turn improves the estimation of the linear decision rule based on a semiparametric working model. Specifically, the proposed semi-supervised (SS) method is a two-step estimation procedure. 
In the first step, we develop a kernel estimator of the contrast function based on the labeled data. In the second step, the contrast function is imputed for unlabeled data based on their covariate information, and then a linear decision rule is estimated based on the imputed contrast function. 

The article is organized as follows. Section \ref{SS_EMR} provides an overview for optimal treatment decision using EMR data. The framework and assumptions are also presented here. An estimation procedure for the optimal linear treatment regime using the labeled data only is introduced in Section \ref{TR-OLS}. The proposed semi-supervised learning method using both labeled and unlabeled data is introduced in Section \ref{Q-Learn}. Simulation studies are conducted in Section \ref{Sims} to compare the empirical performance of the methods introduced in Sections \ref{TR-OLS} and \ref{Q-Learn}. An application to an EMR data set for studying patients in the intensive care unit (ICU) undergoing a hypotensive episode is given in Section \ref{EMR_Analysis}. The paper is concluded with discussions. All the proofs are provided in the Appendix. 

\section{Optimal Treatment Decision with EMR Data} \label{SS_EMR}

When estimating optimal treatment regime using EMR data, a challenge arises from that  a large portion of outcomes of interest and/or treatment information may not be available. For example, \cite{lee2012interrogating} performed a retrospective study on patients in MIMIC-III that underwent a single hypotensive episode during their ICU stay. The goal is to examine outcomes of patients receiving the IV fluid resuscitation or vasoactive therapy. In this study, the treatment information is not recorded for many subjects, despite one of the treatments was typically used as a first line of therapy against hypotension \citep{kellum2012kidney}. In addition to the missingness in the treatment information, one primary outcome of interest, serum creatinine, may also be missing, even though it is a marker for kidney damage, a concern for patients with hypotension. In this work, we will develop a semi-supervised learning method for estimating  optimal treatment regime  when missingness occurs in the response and/or treatment for a large proportion of subjects. 

Next, we present the framework and assumptions for the considered semi-supervised optimal treatment decision problem.  Let $A \in \mathcal{A}$ denote the treatments received by a patient, where $\mathcal{A}$ is the set of possible treatments. Here, we only consider binary treatment options, i.e. $\mathcal{A}= \{0,1\}$. Let $\M{X} \in \mathcal{X} \subseteq \mathbb{R}^{p}$ be a vector of covariates ascertained prior to treatment. 
Let $Y \in \Real$ denote the outcome of interest. Here, a larger value of $Y$ indicates a better outcome. The observed data consist two parts: a set of fully observed data, namely the labeled data, $\mathcal{O} = \Bigl\{ \mathcal{O}_i = (Y_i,A_i,\M{X}_i ): i=1,\ldots,n\Bigr\}$; and a set of partially  observed data, namely the unlabeled data, $\mathcal{U} = \Bigl\{ \M{X}_j : j=n+1,\ldots,n+N\Bigr\}$.  Here, $N$ can be  much larger than  $n$. 
In this work, we assume that the response and treatment information in $\mathcal{U}$ are missing completely at random (MCAR). 

Let $Y^*(a)$ denote the potential outcome that would be observed if a patient were given treatment $a$, where  $a = 1$ or 0. We make the following standard assumptions in causal inference literature \citep{rubin1978bayesian,rosenbaum1983central}: (i) \textit{stable unit treatment value assumption (SUTVA)} $Y_i = A_i Y^*(1) + (1-A_i)Y^*(0)$; (ii) \textit{no unmeasured confounders} $\{Y^{*}(1),Y^{*}(0)\} \perp A \mid \M{X}$;  (iii) \textit{positivity assumption} $0 < \pi(\M{X}) < 1$ for any $\M{X}$, where $\pi(\M{X}) = P(A = 1 \mid \M{X})$ denotes the propensity score. 

A treatment regime is a function that maps the covariates onto the treatment space, $\delta(\M{X}) : \mathcal{X}  \rightarrow \mathcal{A}$. Define $Y^*(\delta) = \delta(\M{x}) Y^*(1) + \{1-\delta(\M{x})\}Y^*(0)$. The optimal treatment regime is defined as $\delta^{\mathit{opt}} = \argmax_{\delta \in \Delta} E\left\{Y^*(\delta) \right\}$, where $\Delta$ is the class of interested treatment regimes.


\section{Optimal Treatment Decision using Labeled Data Only}
\label{TR-OLS}

We assume the following model for the $Q$-function, i.e. 
$$
Q(A,\M{X}) = E(Y \mid \M{X} , A) = \mu(\M{X}) + AC(\M{X}), 
$$
where $\mu(\M{X})$ 
is the baseline mean function and $C(\M{X})$ is the contrast function. Here, both functions $\mu(\M{X})$  and $C(\M{X})$ are unspecified. Under the assumptions (i) and (ii), the optimal treatment regime is given by $\delta^{\mathit{opt}}(\M{X}) = I\left\{C(\M{X}) > 0\right\}$. 

In this work, we are interested in a class of linear treatment regime $\delta(\M{X}) = I(\bm{\beta}^{\prime}\widetilde{\M{X}} > 0)$, where $\widetilde{\M{X}} = \left( 1  , \M{X}^{\prime} \right)^{\prime}$, because they are easily interpretable and implementable in practical applications. In other words, we posit the following working model: $ E\left( Y \mid \mathbf{X} , A \right) = \mu(\mathbf{X}) + A(\bm{\beta}^{\prime}\widetilde{\mathbf{X}})$. To estimate the parameters $\bm{\beta}$ in the working model, we adopt a similar approach in \cite{Songetal2017} based on transformed responses. Specifically, define
$$\widetilde{Y}_i = \frac{Y_i\{A_i - \hat{\pi}(\M{X}_i)\}} {\hat{\pi}(\M{X}_i)\{1 - \hat{\pi}(\M{X}_i)\}},\qquad i=1,\cdots,n,$$
where $\hat{\pi}(\cdot)$ is a consistent estimate for the propensity score. For example, we can posit a logistic regression model for the propensity score and $\hat{\pi}$ can be obtained by the maximum likelihood estimation using the labeled data. 

When $\hat{\pi}$ is replaced by the true propensity score in $\tilde{Y}_i$, it can be shown that $E(\tilde{Y}_i | \M{X}_i) = C(\M{X}_i)$.   
Therefore, an estimator for $\bm{\beta}$ in the working model can be obtained by 
\begin{equation}
\widehat{\bm{\beta}}_{TR} = \argmin_{\beta} \sum_{i=1}^n \left(\widetilde{Y_i} -  \bm{\beta}^{\prime}\widetilde{\mathbf{X}}_i \right)^{2}. 
\end{equation}
Define $\bm{\beta}^* = \arg\min_{\bm\beta} E\{C(\M{X}_i)-\bm\beta'\widetilde{\M{X}}_i\}^2$. 
Following the similar arguments in \cite{Luetal2013}, it can be shown that 
\begin{equation}
\sqrt{n} \left( \widehat{\V{\beta}}_{TR} - \V{\beta}^* \right) = n^{-1/2} \sum_{i=1}^{n} \Psi_{TR}\left( \mathcal{O}_i \right) + o_p( 1), 
\end{equation}
where $\Psi_{TR}\left( \mathcal{O}_i \right)$ are i.i.d. random vectors with mean zero. It follows that $\sqrt{n} \left( \widehat{\V{\beta}}_{TR} - \V{\beta}^* \right)$ converges in distribution to a normal random vector with mean zero and variance $\mathcal{V}_{\V{\beta}_{TR}} = E \left\{ \Psi_{TR} \left( \mathcal{O}_i \right) \Psi_{TR}^{\prime} \left( \mathcal{O}_i \right) \right\}$. Here, $\mathcal{V}_{\V{\beta}_{TR}}$ can be consistently estimated by  
$$
\widehat{\mathcal{V}}_{\V{\beta}_{TR}} = n^{-1} \sum_{i=1}^n \widehat{\Psi}_{TR} \left( \mathcal{O}_i \right) \widehat{\Psi}_{TR}^{\prime} \left( \mathcal{O}_i \right),
$$ 
where $\widehat{\Psi}_{TR} \left( \mathcal{O}_i \right)$ is a consistent estimate of $\Psi_{TR} \left( \mathcal{O}_i \right)$ that can be obtained by the plug-in method. 

\section{Semi-Supervised Learning for Optimal Treatment Decision} \label{Q-Learn}



The proposed semi-supervised learning consists of two steps: (i) compute the kernel estimators of $Q\left(\M{x}, a \right)$ for $ a = 0, 1$ based on labeled data; (ii) impute contrast functions, $ C(\M{X}_j) = Q \left( \M{X}_j, 1 \right) - Q \left( \M{X}_j, 0 \right)$, $j = n+1,...,N$, for unlabeled data, and regress the imputed contrast functions on $\M{X}_j$'s to obtain a semi-supervised estimator for $\V{\beta}$. 

\subsection{Nonparametric Imputation of the $Q$-Functions}
\label{NP-SS-est}

The kernel estimators of $Q(\M{x},a)$ are given by 
\begin{equation} \label{eq_ks1}
\widehat{Q}^{(np)}(\V{x},1) = \dfrac{h^{-p}\sum_{i = 1}^{n} W\left( \frac{\mathbf{x - X_i}}{h} \right) A_i Y_i}{h^{-p}\sum_{i = 1}^{n} W\left( \frac{\mathbf{x - X_i}}{h} \right) A_i}
\end{equation}
and,
\begin{equation} \label{eq_ks0}
\widehat{Q}^{(np)}(\V{x},0) = \dfrac{h^{-p}\sum_{i = 1}^{n} W\left( \frac{\mathbf{x - X_i}}{h} \right) (1 - A_i) Y_i}{h^{-p}\sum_{i = 1}^{n} W\left( \frac{\mathbf{x - X_i}}{h} \right) (1 - A_i) },
\end{equation}
where $W(\cdot): \mathbb{R}^{p} \rightarrow \mathbb{R}$ is a suitable kernel function and $h$ is a bandwidth that converges to zero as $n$ goes to infinity. Define $\widehat{C}^{(np)}(\V{x}) = \widehat{Q}^{(np)}(\V{x},1) - \widehat{Q}^{(np)}(\V{x},0)$.
\ 
The estimate of $\V{\beta}$ based on imputed contrast functions for unlabeled data is given by
\begin{equation} \label{NP_OLS}
\widehat{\bm{\beta}}_{np} = \argmin_{\beta} \sum_{j=n+1}^{n+N} \left\{\widehat{C}^{(np)}(\M{X}_j) -  \bm{\beta}^{\prime}\widetilde{\mathbf{X}}_j \right\}^{2}.
\end{equation}
The corresponding linear treatment regime is $\widehat{\delta}_{np}(\M{X}) = I(\widehat{\V{\beta}}_{np}^{\prime}\widetilde{\M{X}} > 0)$. 

Next, we establish the asymptotic distribution of $\widehat{\V{\beta}}_{np}$. The subsequent assumptions are needed for Theorem \ref{thm1} given below. As in the literature for kernel estimation \citep{fan1992design,hansen2008uniform,newey1994kernel,chakrabortty2018efficient}, we assume the following conditions. 
\begin{itemize}
\item[(C1)] The $Q$-functions $Q(\V{x},a)$, the density function $f(\V{X})$ of $\M{X}$ and the propensity score $\pi(\V{x})$ are $r$ times continuously differentiable with bounded $r^{th}$ derivatives on $\V{X} \in \mathcal{X}$. 
\item[(C2)] The function $W(\cdot)$ is a symmetric $r$-th order kernel for some integer $r \geq 2$. In addition, $W(\cdot)$ is Lipschitz continuous and has bounded support. 
\item[(C3)] \label{as_expect_bound} $E(\lvert AY \rvert^s) < \infty $ and $E(\lvert (1-A)Y \rvert^s) < \infty $ for some $s > 2$. 
\item[(C4)] $E(\lvert A Y \rvert^s \mid \M{X} = \V{x}) f(\V{x}) = E(\lvert Y \rvert^s \mid \M{X} = \V{x}) \pi(\V{x}) f(\V{x})$ and $f(\V{x})$ are bounded on $\mathcal{X}$ and \label{as_inf} $\mathrm{inf}_{\V{x} \in \mathcal{X}} \ f(\V{x}) > 0 $.
\end{itemize}

\begin{theorem} \label{thm1}
Suppose $n^{1/2}h^r \rightarrow 0$ and $\sqrt{\frac{\log n}{nh^p}} \rightarrow 0$ as $h \rightarrow 0$, $n, N \rightarrow \infty$ and $n/N \rightarrow 0$. Then, under Conditions (C1)-(C4), we have
\begin{equation*}
n^{1/2} \left( \widehat{\V{\beta}}_{np} - \V{\beta}^*\right) = n^{-1/2} \sum_{i=1}^{n} \Psi_{np}(\mathcal{O}_i) + o_p( 1), 
\end{equation*}
where $\Psi_{np}(\mathcal{O}_i) = \left\lbrace \frac{A_i}{\pi(\M{X}_i)} - \frac{1-A_i}{1-\pi(\M{X}_i)} \right\rbrace \Lambda^{-1} \widetilde{\M{X}}_i \left\{Y_i - Q(\M{X}_i,A_i) \right\}$ and $\Lambda = E(\widetilde{\M{X}}_j\widetilde{\M{X}}_j')$. 
 
\end{theorem}

Theorem \ref{thm1} implies that $n^{1/2}(\widehat{\V{\beta}}_{np} - \V{\beta}^*)$ converges in distribution to a multivariate normal $\mathit{N}_{p+1} (0 , \mathcal{V}_{\V{\beta}_{np}})$
with $\mathcal{V}_{\V{\beta}_{np}} = E \left\{\Psi_{np} \left( \mathcal{O}_i \right) \Psi_{np} ^{\prime}\left( \mathcal{O}_i \right) \right\}$. The asymptotic variance matrix $\mathcal{V}_{\V{\beta}_{np}}$ can be consistently estimated by the usual plug-in method. 


As discussed in \cite{fan1992design}, kernel estimates may suffer finite sample bias especially when sample size is small. This in turn will cause relatively large finite sample bias in $\widehat{\V{\beta}}_{np}$ based on imputed contrast functions. In the next section, we develop a similar cross-validation technique to correct the finite sample bias in kernel estimation as in \cite{chakrabortty2018efficient}. In addition, when the number of covariates is relatively large, some other nonparametric regression methods, such as random forest, may be used for estimating the $Q$ functions since kernel estimation may suffer the curse of dimensionality.         

\subsection{Semi-Nonparametric Imputation of the Q-Functions} \label{SS_section}

Consider a $\mathcal{K}$-fold cross-validation. Let $\mathcal{O}^{(k)}$, $k = 1,..,\mathcal{K}$, denote the corresponding partition. First, we compute the kernel estimators of the $Q$-functions based on data excluding the $k$-th fold, denoted by $\widehat{Q}_{-k}^{np}(\V{x},a)$. Next, we conduct a linear refitting step to correct the bias of $\widehat{Q}^{(np)}(\V{x},a)$. Specifically, define  
\begin{equation}
\widehat{\V{\theta}}_1 = \argmin_{\V{\theta}_1} \sum_{k=1}^{\mathcal{K}} \sum_{i \in \mathcal{O}^{(k)}} \frac{A_i}{\hat{\pi}(\M{X}_i)} \left\{ Y_i - \widehat{Q}_{-k}^{(np)}\left( \mathbf{X}_i, A_i \right) - \V{\theta}_1^{\prime}\widetilde{\mathbf{X}}_i \right
\}^2,
\end{equation}
and
\begin{equation}
\widehat{\V{\theta}}_0 = \argmin_{\V{\theta}_0} \sum_{k=1}^{\mathcal{K}} \sum_{i \in \mathcal{O}^{(k)}} \frac{1-A_i}{1-\hat{\pi}(\M{X}_i)} \left\{ Y_i - \widehat{Q}_{-k}^{(np)}\left( \mathbf{X}_i, A_i \right) - \V{\theta}_0^{\prime}\widetilde{\mathbf{X}}_i \right\}^2.
\end{equation}
The semi-nonparametric estimators of the $Q$-functions are given by 
\begin{equation}
\widehat{Q}^{(SS)} ( \V{x},a , \widehat{\V{\theta}}_a )  = \dfrac{1}{\mathcal{K}} \sum_{k=1}^{\mathcal{K}} \widehat{Q}_{-k}^{(np)}\left( \V{x}, a \right)  +  \widehat{\V{\theta}}_a^{\prime}\widetilde{\V{x}},
\end{equation}
$\mathrm{for} \ a \in \left\lbrace  0,1 \right\rbrace$. Define $\widehat{C}^{(SS)}(\V{x}) = \widehat{Q}^{(SS)} ( \V{x}, 1 , \widehat{\V{\theta}}_1 ) - \widehat{Q}^{(SS)} ( \V{x}, 0 , \widehat{\V{\theta}}_0 )$. A least squares estimate $\widehat{\V{\beta}}_{SS}$ can be obtained as in \eqref{NP_OLS} based on imputed contrast functions $\widehat{C}^{(SS)}(\M{X}_j)$, $j=n+1,...,N$. The corresponding linear treatment regime is $\widehat{\delta}_{SS}(\M{X}) = I(\widehat{\V{\beta}}_{SS}^{\prime}\widetilde{\M{X}} > 0)$.

\begin{theorem} \label{thm2}
Under the same conditions assumed in Theorem \ref{thm1}, we have
\begin{equation*}
n^{1/2} \left( \widehat{\V{\beta}}_{SS} - \V{\beta}^* \right) = n^{-1/2} \sum_{i=1}^{n} \Psi_{SS}(\mathcal{O}_i) + o_p( 1), 
\end{equation*}
where $\Psi_{SS}(\mathcal{O}_i) = \left\lbrace \frac{A_i}{\pi(\M{X}_i)} - \frac{1-A_i}{1-\pi(\M{X}_i)} \right\rbrace \Lambda^{-1} \widetilde{\M{X}}_i \left\{ Y_i - Q^{(SS)}(\M{X}_i, A_i, \V{\theta}_A) \right\}$ and $Q^{(SS)} \left( \V{x},a, \V{\theta}_a \right) = Q \left( \V{x},a \right) + \V{\theta}_a^{\prime}\V{x}$. Here $\V{\theta}_a$ is the limit of $\widehat{\V{\theta}}_a$, $a= 0, 1$.
\end{theorem}


The asymptotic variance of $\widehat{\V{\beta}}_{SS}$ can be estimated by the usual plug-in method. To reduce the bias in the variance estimation, a double cross-validation method as proposed in \cite{chakrabortty2018efficient} can be adopted.  The proof of Theorems \ref{thm1} and \ref{thm2} are outlined in the Appendix.

\section{Simulation Analysis} 
\label{Sims}
We conducted simulation studies to examine the finite-sample performance of the estimators introduced in Section \ref{TR-OLS} (denoted by TR) and in Section 4.2 (denoted by SS). Our simulation study design considers varying degrees of model misspecification in $C(\M{X})$ to better understand the advantages of SS when estimating $\delta_{\V{\beta}}^{opt}$. During each simulation replication, the propensity score will be estimated to better reflect the analysis of an observation study with EMR data. In the implementation of our proposed SS estimator, a Gaussian kernel was used for $W(\cdot)$ and the bandwidth parameter $h$ was chosen using cross-validation. In addition we set $\mathcal{K}=5$. We present the percent of correct decisions (PCD) and value function to illustrate the benefits of semi-supervised prediction on assessing $\delta_{\V{\beta}}^{\mathit{opt}}$. Furthermore, we study the bias, empirical standard deviation (SD), estimated standard error (SE), and the relative efficiency of $\widehat{\V{\beta}}_{SS}$ with respect to $\widehat{\V{\beta}}_{TR}$.
\subsection{Data Generation}
In this simulation, $n$ is the amount of subjects in $\mathcal{O}$ and $N$ is the amount of subjects in $\mathcal{U}$. The $N$ unobserved responses were simulated to be missing completely at random (MCAR), where the response and treatment were not observed. Let $\mathbf{X} \sim N\left(\V{0}_p,I_{p} \right)$, where $\M{X}$ is contained in $[-5,5]^p$ to ensure it is in a compact set, and $\mathcal{A} \in \left\lbrace0,1 \right\rbrace$. The following models were tested with $p=2$, $n = 500$ and $N = 5000$. We examined three choices for $C(\M{X})$, 

\begin{itemize}
	\item \emph{Model 1 (Linear)}: $\mathbf{Y = \mu(\mathbf{X}) + \mathbf{A}\left( \bm{\eta}^{\prime} \mathbf{X} \right)  + \epsilon}$
	\item \emph{Model 2 (Cubic)}: $\mathbf{Y = \mu(\mathbf{X}) + \mathbf{A}\left(\bm{\eta}^{\prime} \mathbf{X} \right)^3   + \epsilon}$ 
    \item \emph{Model 3 (Sine)}: $\mathbf{Y = \mu(\mathbf{X}) + \mathbf{A} \left( \textrm{sin}(\bm{\eta}^{\prime} \mathbf{X} ) \right) + \epsilon}$,
\end{itemize}

\noindent where $\bm{\eta} = \mathbf{1_{2}}$ for Models 1 and 3, $\bm{\eta} = \left( 0.3,0.6 \right)^\prime$ for Model 2, and $\mathbf{\epsilon} \sim \mathcal{N}(0 , 1)$. Two baseline functions were studied. The first one considered a situation where the baseline effect for subjects is small and the second function considered a larger baseline effect,
\begin{enumerate}
	\item $\mu(\mathbf{X})=\left( \V{\omega}^{\prime} \mathbf{X} \right)^3 $,
	\item $\mu(\mathbf{X})= \left( \bm{\alpha}^{\prime}\mathbf{X} \right) \left( 1+ \V{\omega}^{\prime} \mathbf{X} \right)$.
\end{enumerate}

We studied two different baseline functions to understand the flexibility of SS in estimating $C(\M{X})$. The models were studied with $\V{\omega} = (0.5,0.5)^{\prime}$ and $\V{\alpha} = (0.75,0.75)^\prime$. We simulated a propensity score of $\pi(\mathbf{X}) = logit(0.5X_1 - 0.5 X_2 )$, and an allocation of treatment as $\mathbf{A} \sim bernoulli(p=\pi(\mathbf{X}))$. The propensity scores were generated to simulate observational data where the allocation of treatment depended on a patient's health status information. The true values, $\bm{\beta}^*$, were estimated by simulating a fully observed Monte Carlo data set of size 500,000.

\subsection{Simulation Results} 
\label{Sim_results}
The results of interest were percent of correct decisions (PCD), value function (V), and component-wise relative efficiency (RE) over 500 replications. The component-wise RE for each estimator was calculated as $\sum_{i=1}^{500} \lVert \widehat{\bm{\beta}}_{TR,i,j} - \bm{\beta}^*_{j} \rVert^2 / \sum_{k=1}^{500} \lVert \widehat{\bm{\beta}}_{SS,j} - \bm{\beta}^*_{j} \rVert^2$, where $j=1, \cdots, p+1$. During each replication, new vectors $\mathbf{Y}$, $\mathbf{A}$ and matrix $\mathbf{X}$ were simulated. The percent of correct decisions (PCD) was calculated for each method, 
\begin{equation*}
PCD_i = 1 -  \dfrac{1}{5500}\sum_{k=1}^{5500} \left| I ( \widehat{\bm{\beta}^{\prime}} \widetilde{\M{X}}_k > 0 )  - I ( \bm{\beta}^{*\prime} \widetilde{\M{X}}_k > 0 ) \right|, 
\end{equation*}
and then averaged over all 500 simulations, $PCD =  (1/500) \sum_{i=1}^{500} PCD_i$. The true value function was calculated with the Monte Carlo data set of sample size 500,000 as, 
\begin{equation*}
V_0 = \dfrac{1}{500,000}\sum_{m=1}^{500,000}\left\lbrace \mu(\mathbf{X}_m) + I(C(\M{X}_m)>0) C(\M{X}_m) \right\rbrace.
\end{equation*}

\noindent Estimation of the value function using TR or SS was calculated as,
\begin{equation*}
\widehat{V}_i = \dfrac{1}{500,000}\sum_{m=1}^{500,000}\left\lbrace \mu(\M{X}_m) +\widehat{\delta}_{\V{\beta}}^{opt} C(\M{X}_m) \right\rbrace,
\end{equation*}

\noindent on the same Monte Carlo data set and then averaged over 500 replications. 
\\
\indent Table~\ref{table:VF_RE_PCD_500} summarized the value function and PCD results under all 6 settings. Across all settings, we observed an improvement in the average value function, average PCD, and their respective SD while using the SS method compared to TR. Table~\ref{table:500coefs} presented the mean bias, standatrd deviation (SD), average of estimated standard errors (SE), and coverage probabilities of the 95$\%$ Wald-type confidence intervals of the estimates over 500 replications. In addition, the relative efficiency (RE) of the SS estimators compared with the TR estimator was also reported. Based on the results, we observed that both estimators are nearly unbiased, but the the SS estimators are more efficient that the TR estimator. The RE ranged from 2.13 to 5.37.



\section{Application to an EMR Study}
\label{EMR_Analysis}
We apply our proposed method, SS, to an EMR study on patients undergoing a hypotensive episode in the ICU within the MIMIC-III database. It is important to treat hypotensive episodes (HE) in ICU patients to minimize end-organ damage. A marker of end-organ damage is an increase in serum creatinine post hypotensive episode \citep{lehman2010hypotension}. Specifically, the outcome of interest considered in our study is the difference between a post-episode serum creatinine measurement taken no more than 72 hours after the HE and a baseline serum creatinine measurement taken no earlier than 24 hours before the HE. Two available treatments for HE's include IV fluid resuscitation and vasoactive therapy \citep{lee2012interrogating}. Our objective is to determine the optimal initial treatment for each individual between IV fluid resuscitation and vasopressor interventions to minimize any increase in serum creatinine 72 hours post-hypotensive episode. 
\\
\indent \cite{lee2012interrogating} defines the initiation of an HE as the occurrence of two successive mean arterial pressure (MAP) measurement values $>$ 60 \textit{mm Hg}, followed by two MAP measurements $\leq$ 60 \textit{mm Hg}. The end of an HE is characterized by two successive MAP measurements $\leq$ 60 \textit{mm Hg}, followed by two successive MAP measurements $>$ 60 \textit{mm Hg}. The study cohort includes subjects greater than or equal to 15 years old who experienced a single hypotensive episodes during their ICU stay, which amounts to 3,372 subjects.
\\
\indent To formulate this problem into the OTD framework, we must define $Y$ as the negative of the difference between post-HE and pre-HE serum creatinine measurements. The treatment space is defined as $A = 1$ if the patient was given vasopressor treatment, and $A=0$ if they received IV fluid resuscitation. Vasopressor treatments include dobutamine, dopamine, epinephrine, norepinephrine, phenylephrine, milrinone or vasopressin during the HE. IV fluid treatment is defined as receiving at least one infusion of colloids, or isotonic crystalloids of at least 250 ml \citep{lee2012interrogating}. The predictors, $\M{X}$, include normalized versions of mean MAP and mean heart rate in the 3 hour window immediately before the HE onset. A total of 945 subjects received either IV fluid resuscitation or vasopressor treatment as their sole treatment. The number of patients untreated with solely IV fluid rescuitation or vasoactive therapy and/or missing response information resulted in 2,427 subjects. For this data analysis, we are focusing on the semi-supervised setting with a small set of labeled data. Therefore we randomly choose $n=300$ out of 945 subjects to create the fully observed set, $\mathcal{O}$, and place the remaining 3,072 subjects into $\mathcal{U}$. Under such a sampling scheme, the missing completely at random assumption is held. 


\indent Propensity scores were modeled with a logistic regression model that included the following covariates: baseline creatinine, age, gender, indicator function for patients that received surgical services, log transformation of the  elixhauser comorbidity score (method of categorizing comorbidity of patients based on their ICD-9 codes) \citep{van2009modification}, log transformation of simplified acute physiologic score (SAPS) (scoring system to reflect the risk of death upon admission to ICU) \citep{le1984simplified}, total urine output in the 3 hour window prior to the HE, average mean heart rate in the 3 hour window prior to the HE, and average mean arterial pressure (MAP) in the 3 hour window prior to the HE. We perform TR and SS on the data set to obtain estimates and standard errors for $\V{\beta}$. A Gaussian kernel was chosen for $W(\cdot)$ in Equations \ref{eq_ks1} and \ref{eq_ks0} and $\mathcal{K}=5$ similar to Section \ref{Sims}. Table \ref{table:RealData} presents the point estimates, estimated SE, and p-values for testing null effects. SS and TR estimates are relatively close, but SS estimates have smaller SE than TR estimates suggesting they are more efficient as demonstrated in simulations.

In Table~\ref{table:allocation}, we demonstrate treatment allocation given by both methods on all subjects in the data set. The OTD estimated by SS and TR produce similar treatment decisions for the vast majority of subjects. SS assigns more patients to treatment with IV fluids than TR and is less likely to assign vasoactive therapy.  



\section{Discussion}
\label{Future Work}

 We proposed a new semi-supervised learning method for estimating optimal treatment regime using both labeled and unlabeled data from EMR. In this work, we focused on binary treatments, but the proposed method can be easily extended to accommodate multiple treatments. In addition, we considered a semi-parametric working model with linear treatment effects for its simplicity. The estimated linear treatment regime may not give the highest value function among the class of linear decision rules since it is obtained based on a regression method. Alternatively, we may consider an estimated linear treatment regime that maximizes the estimated value function or equivalently the estimated contrast function. However, the resulting estimator of $\V{\beta}$ has a slower cubic convergence rate and its associated inference will become more challenging. Moreover, the proposed method can be generalized to accommodate nonlinear treatment effects using spline representation techniques.        
 
 The proposed method is built on kernel estimation for imputing the contrast functions of unlabeled data, which can not handle high-dimensional predictors. To incorporate high-dimensional predictors, some dimension reduction techniques, such as principal component analysis or sliced inverse regression, can be utilized to reduce the dimension of covariates into a lower-dimensional subspace, and then kernel estimators can be applied as in \cite{chakrabortty2018efficient}. 
 Moreover, in this paper we only consider the MCAR setting, but the proposed semi-supervised learning method can be easily extended to the case when $Y$ and $A$ are missing at random (MAR) by using the inverse missing probability weighted estimation. 

The proposed semi-supervised learning method only uses variables obtained from structured EMR data. MIMIC-III also contains unstructured data through progress reports and discharge summaries recorded during a patient's hospital stay. Clinical text data holds a great deal of valuable patient information that can be used for building the optimal treatment regime. Clinical natural language processing tools, such as cTakes, can be used to extract information from clinical texts, such as patients' disease status, symptoms, prescribed treatments, and family history \citep{savova2010mayo}. 
How to incorporate these information in the optimal treatment decision is an interesting question that warrants future investigation. 

\appendix
\label{appendix1}

\section{Proof of Theorem 1.}
Let $\widehat{C}(\V{X}) = \widehat{Q}^{(np)}(\V{X},1) - \widehat{Q}^{(np)}(\V{X},0)$. We have
\begin{align*}
 \widehat{\V{\beta}}_{np} - \V{\beta}^*  &= \Lambda^{-1} \left[ N^{-1} \sum_{j=n+1}^{n+N} \widetilde{\M{X}}_j \left\lbrace \widehat{C}(\M{X}_j) - {\V{\beta}^*}^{\prime} \widetilde{\M{X}}_j  \right\rbrace \right] \\
&= \Lambda^{-1} \left[ N^{-1} \sum_{j=n+1}^{n+N} \widetilde{\M{X}}_j \left\lbrace \widehat{C}(\M{X}_j) - C(\M{X}_j) \right\rbrace \right] +  \Lambda^{-1} \left[ N^{-1} \sum_{j=n+1}^{n+N} \widetilde{\M{X}}_j \left\lbrace C(\M{X}_j) - \V{\beta}^{\prime} \widetilde{\M{X}}_j \right\rbrace \right] \\
&= \Lambda^{-1} E \left[ \widetilde{\M{X}} \left\lbrace \widehat{C}(\M{X}) - C(\M{X}) \right\rbrace \right] + O_p(N^{-1/2}).
\end{align*}

\noindent The first step follows from the normal equations. The last step is due to the fact 
$$
\Lambda^{-1} \left[ N^{-1} \sum_{j=n+1}^{n+N} \widetilde{\M{X}}_j \left\lbrace \widehat{C}(\M{X}_j) - C(\M{X}_j) \right\rbrace \right] = \Lambda^{-1} E \left[ \widetilde{\M{X}} \left\lbrace \widehat{C}(\M{X}) - C(\M{X}) \right\rbrace \right] + o_p(1)
$$ 
by standard arguments involving the weak law of large numbers. According to the central limit theorem $N^{-1/2} \left[ N^{-1/2} \sum_{j=n+1}^{N} \Lambda^{-1} \left\lbrace C(\M{X})_j - \V{\beta}^{\prime} \widetilde{\M{X}}_j \right\rbrace \right] = O_p(N^{-1/2})$. Multiplying both sides by $n^{1/2}$ we have
\begin{eqnarray}
n^{1/2} \left( \widehat{\V{\beta}}_{np} - \V{\beta} \right) &=& n^{1/2} \Lambda^{-1} E \left[ \widetilde{\M{X}} \left\lbrace \widehat{C}(\V{X}) - C(\M{X}) \right\rbrace \right] + O_p \left( (n/N)^{\frac{1}{2}} \right) \\
&=& n^{1/2} \Lambda^{-1} E \left[ \widetilde{\M{X}} \left\lbrace \widehat{Q}^{(np)}(\M{X},1) - Q^{(np)}(\M{X},1) \right\rbrace \right] \\
&-& n^{1/2} \Lambda^{-1} E \left[ \widetilde{\M{X}} \left\lbrace \widehat{Q}^{(np)}(\M{X},0) - Q^{(np)}(\M{X},0) \right\rbrace \right] + O_p \left( (n/N)^{\frac{1}{2}} \right).
\end{eqnarray}

Note that $n/N \rightarrow 0$ implying $O_p \left( (n/N)^{\frac{1}{2}} \right) \equiv o_p(1)$. Next, let $\tau(\M{x}) = \pi(\M{x})f(\M{x})$ and $\widehat{\tau}(\M{x}) = \frac{1}{nh^p} \sum_{i=1}^n A_i W_h(\M{X}_i - \M{x})$, where $W_h(\M{X}_i - \M{x}) = W(\frac{\M{X}_i - \M{x}}{h})$. Let's rewrite $E \left[ \widetilde{\M{X}} \left\lbrace \widehat{Q}^{(np)}(\M{X},1) - Q^{(np)}(\M{X},1) \right\rbrace \right]$ as
\begin{align} \label{def_alphax}
&= E \left\lbrace \widetilde{\M{X}} \left( \dfrac{\frac{1}{nh^p}\sum_{i=1}^n A_i W_h(\M{X}_i - \M{x}) \left\lbrace Y_i - Q^{(np)}(\M{x},1) \right\rbrace}{\tau(\M{x})} \right) \right\rbrace \\
&+ E \left\lbrace \widetilde{\M{X}} \left( \widehat{Q}^{(np)}(\M{X},1) - Q^{(np)}(\M{X},1) \right) \left\lbrace \dfrac{\tau(\M{x}) - \widehat{\tau}(\M{x})}{\tau(\M{x})} \right\rbrace \right\rbrace = H_{n,1}^{(1)} + H_{n,2}^{(1)}.
\end{align}

Then $H_{n,1}^{(1)}$ is equivalent to:

\begin{eqnarray*}
&=& \frac{1}{nh^p} \sum_{i=1}^n A_i \int \widetilde{\M{X}}_i \left\lbrace Y_i - Q^{(np)}(\M{X},1) \right\rbrace \dfrac{W_h(\M{X} - \M{X}_i)}{\tau(\M{x})} f(\M{x})d\M{x} \label{eqn:1} \\
&=& \frac{1}{nh^p} \sum_{i=1}^n A_i \int \widetilde{\M{X}}_i \left\lbrace Y_i - Q^{(np)}(\M{X},1) \right\rbrace \dfrac{W_h(\M{X} - \M{X}_i)}{\pi(\M{x})} d\M{x} \label{eqn:2} \\
&=& \frac{1}{n} \sum_{i=1}^n A_i \int \left( \widetilde{\M{X}}_i + h \V{t}_i \right) \left\lbrace Y_i - Q^{(np)}(\M{X}_i + h \V{t}_i,1) \right\rbrace \dfrac{W(\V{t}_i)}{\pi\left( \M{X}_i + h \V{t}_i \right)} d \V{t}_i. \label{eqn:3}
\end{eqnarray*}

By  assumptions (C1) and (C2), and Taylor expansion of $H_{n,1}^{(1)}$ in $h \V{t}_i$ for sufficiently small $h$, it leads to
\begin{equation*}
H_{n,1}^{(1)} = \frac{1}{n} \sum_{i=1}^n \dfrac{A_i}{\pi\left( \M{X}_i \right)} \widetilde{\M{X}}_i \left\lbrace Y_i - Q^{(np)}(\M{X}_i,1) \right\rbrace + O_p(h^r).
\end{equation*}
Since $n^{1/2}h^r \rightarrow 0$ as $n \rightarrow \infty$,
\begin{equation}
n^{1/2} \Lambda^{-1} H_{n,1}^{(1)} = \widetilde{{H}}_{n,1}^{(1)} = n^{-1/2} \sum_{i=1}^n \dfrac{A_i}{\pi\left( \M{X}_i \right)} \Lambda^{-1} \widetilde{\M{X}}_i \left\lbrace Y_i - Q^{(np)}(\M{X}_i,1) \right\rbrace + o_p(1).
\end{equation}
Let $q(\M{X}) = \widehat{Q}^{(np)}(\M{X},1) - Q^{(np)}(\M{X},1)$, $l(\M{X}) = \dfrac{\tau(\M{x}) - \widehat{\tau}(\M{x})}{\tau(\M{x})}= 1 - \dfrac{ \widehat{\tau}(\M{x})}{\tau(\M{x})}$. Given that $\M{X}$ is bounded, $\sqrt{\frac{\ln n}{nh^p}} \rightarrow 0$ as $n \rightarrow \infty$, as well as assumptions (C1), (C3), and (C4), \textit{Lemma B.1} in \cite{newey1994kernel}, it can be shown that $\mathrm{sup}_{\V{x} \in \mathcal{X}} \lvert \widehat{\tau}(\M{x}) - \tau(\M{X}) \rvert = o_p(1)$. It also holds that $\mathrm{sup}_{\V{x} \in \mathcal{X}} \lvert q(\M{x}) \rvert = o_p(1)$ through \textit{Theorem 8} of \cite{hansen2008uniform}. It follows from these results that $H_{n,2}^{(1)} = E \left\lbrace \widetilde{\M{X}} q(\M{X}) l(\M{X}) \right \rbrace$ and,
\begin{eqnarray} \label{sup_bound_expect}
E \left\lbrace \widetilde{\M{X}} q(\M{X}) l(\M{X}) \right \rbrace \leq \mathrm{sup}_{\V{x} \in \mathcal{X}} \left \lbrace \norm{\widetilde{\M{X}}} \ \lvert q(\M{x}) \rvert \ \lvert l(\M{x}) \rvert  \right\rbrace = o_p(1).
\end{eqnarray}

Moreover, using similar techniques we can prove,
\begin{eqnarray} \label{eqn_18_prf1}
n^{1/2}\Lambda^{-1} E \left[ \widetilde{\M{X}} \left\lbrace \widehat{Q}^{(np)}(\M{X},0) - Q^{(np)}(\M{X},0) \right\rbrace \right] = \widetilde{H}_{n,1}^{(0)} + o_p(1).
\end{eqnarray}
This leads to 
$$n^{1/2} \Lambda^{-1} E \left[ \widetilde{\M{X}} \left\lbrace \widehat{C}(\V{X}) - C(\M{X}) \right\rbrace \right] = \widetilde{H}_{n,1}^{(1)} - \widetilde{H}_{n,1}^{(0)} + o_p(1) = n^{-1/2} \sum_{i=1}^{n} \Psi_{np}( \M{O}_i) + o_p( 1 ). $$ 
It approves the asymptotic results established in Theorem 1.

\section{Proof of Theorem 2.} \label{Appendix_proof2}
The proof of Theorem 2 follows similarly to that of Theorem \ref{thm1}. Let $\widehat{C}^{(SS)}(\M{X}) = \widehat{Q}^{(SS)} ( \M{x},1, \widehat{\V{\theta}}_1 ) - \widehat{Q}^{(SS)}(\M{X},0, \widehat{\V{\theta}}_0)$. We have
\begin{eqnarray}
n^{1/2}\left( \widehat{\V{\beta}}_{SS} - \V{\beta} \right) &=& n^{1/2} \Lambda^{-1} E \left[ \widetilde{\M{X}} \left\lbrace \widehat{C}(\V{X}) - C(\M{X}) \right\rbrace \right] + o_p \left( 1 \right) \label{Eqn_19} \\
&=& n^{1/2} \Lambda^{-1} E \left[ \widetilde{\M{X}} \left\lbrace  \widehat{Q}^{(SS)} ( \M{x},1, \widehat{\V{\theta}}_1 ) -  Q^{(SS)} ( \M{x},1, \V{\theta}_1 ) \right\rbrace \right] \label{Eqn_20} \\
&-& n^{1/2} \Lambda^{-1} E \left[ \widetilde{\M{X}} \left\lbrace \widehat{Q}^{(SS)} ( \M{x},0, \widehat{\V{\theta}}_0 ) - Q^{(SS)} ( \M{x},0, \V{\theta}_0 ) \right\rbrace \right] + o_p \left( 1 \right). \label{Eqn_21}
\end{eqnarray}
Next, recall that $$\widehat{Q}^{(SS)} ( \V{x},a, \widehat{\V{\theta}}_a ) = \dfrac{1}{\mathcal{K}} \sum_{k=1}^{\mathcal{K}} \widehat{Q}_k^{(np)}\left( \V{x}, a \right)  +  \widehat{\V{\theta}}_a^{\prime}\widetilde{\V{x}},$$ 
and 
$$Q^{(SS)} ( \V{x},a, \V{\theta}_a ) = Q^{(np)}\left( \V{x}, a \right)  + \V{\theta}_a^{\prime}\widetilde{\V{x}}.$$ 
This allows us to rewrite Equation (\ref{Eqn_20}) in the following way
\begin{equation} \label{Eqn_22}
    n^{1/2} \Lambda^{-1} \left\lbrace \frac{1}{N} \sum_{j=n+1}^N \widetilde{\M{X}}_j \left\lbrace  \dfrac{1}{\mathcal{K}} \sum_{k=1}^{\mathcal{K}} \widehat{Q}_k^{(np)}\left( \M{x}_j, a \right) + \widehat{\V{\theta}}_a^{\prime}\widetilde{\M{x}}_j \right \rbrace - Q^{(np)}\left( \M{x}_j, a \right)  + \V{\theta}_a^{\prime}\widetilde{\M{x}}_j  \right\rbrace + o_p(1).
\end{equation}

Equation (\ref{Eqn_21}) can be rewritten analogously. Define
$$\widehat{S}^{\mathcal{K}}_{a}= \dfrac{1}{\mathcal{K}} \sum_{k=1}^{\mathcal{K}} \left\lbrace \dfrac{1}{N} \sum_{j=n+1}^{n+N} \widetilde{\M{X}}_j \left\lbrace \widehat{Q}_k^{(np)} ( \M{x}_j,a) - Q^{(np)} ( \M{x}_j,a) \right\rbrace \right\rbrace.$$
Rearranging the summands in Equation (\ref{Eqn_22}), we can obtain 
\begin{equation}
	 n^{1/2}\left( \widehat{\V{\beta}}_{SS} - \V{\beta} \right) = n^{1/2} \Lambda^{-1} \left \lbrace \Lambda (\widehat{\V{\theta}}_1 - \V{\theta}_1) + \widehat{S}^{\mathcal{K}}_{1} \right \rbrace - n^{1/2} \Lambda^{-1} \left \lbrace \Lambda (\widehat{\V{\theta}}_0 - \V{\theta}_0) + \widehat{S}^{\mathcal{K}}_{0} \right \rbrace + o_p(1).
\end{equation}
Following similar arguments in the proof of \textit{Theorem 3.2} in \cite{chakrabortty2018efficient}, we can approve the asymptotic results established in Theorem \ref{thm2}.   




\vspace*{-8pt}





\vspace*{-8pt}




\bibliographystyle{ECA_jasa}
\bibliography{SSL_OTD}

\newpage

\begin{table}
	\centering 
	\def\~{\hphantom{0}}
	\begin{minipage}{150mm}
	\caption{Treatment decision results: $V_0$, the value of the true optimal treatment rule; $V$, the average of the values of estimated optimal linear treatment rules over 500 replications; PCD, average of percent of correct decisions over 500 replications. Empirical standard deviations are provided in parentheses.} 
	\label{table:VF_RE_PCD_500}
    \begin{tabular*}{\textwidth}{c@{\extracolsep{\fill}}c@{\extracolsep{\fill}}c@{\extracolsep{\fill}}c@{\extracolsep{\fill}}c@{\extracolsep{\fill}}c@{\extracolsep{\fill}}c@{\extracolsep{\fill}}}
		\hline
		\hline
		& & & \multicolumn{2}{c}{TR} & \multicolumn{2}{c}{SS} \\ 
		\cline{4-5} \cline{6-7} \\ [-3pt]		$\mu(\mathbf{X})$ & Model & $V_0$ & V & PCD & V & PCD \\  
		\hline
	    & Linear &0.56 &0.54 (0.04) & 0.92 (0.06) &0.56 (0.01) & 0.96 (0.02) \\
		\footnotesize{$\left( \V{\omega}^{\prime} \M{X} \right)^3$} & Cubic &0.24 &0.22 (0.06) & 0.87 (0.12) &0.24 (0.01) & 0.93 (0.05) \\
		& Sine &0.32 &0.21 (0.12) &0.80 (0.17) &0.26 (0.06) & 0.88 (0.09) \\ [0.5ex]
		\hline
		& Linear &1.31 &1.29 (0.05) & 0.91 (0.06) &1.31 (0.01) & 0.96 (0.02) \\
		\footnotesize{$\left( \V{\alpha}^{\prime} \M{X} \right) \left( 1 + \V{\omega}^{\prime}\M{X} \right)$} & Cubic &0.99 &0.98 (0.05) &0.86 (0.11) &0.99 ($<$0.01) & 0.94 (0.04) \\
		& Sine &1.06 &0.96 (0.11) &0.80 (0.16) &1.02 (0.04) & 0.89 (0.06) \\
		\hline
	\end{tabular*}
  \end{minipage}
  \vspace*{12pt}
\end{table}

\begin{table}
	\centering 
	 \def\~{\hphantom{0}}
		\caption{Parameters estimation results: Bias, the bias of the estimates over 500 replications; SD, empirical standard deviation of the estimates; SE, average of estimated standard errors; CP, empirical coverage probability of Wald-type 95$\%$ confidence intervals; RE, relative efficiency of the SS estimates compared with the TR estimates.} 
	\label{table:500coefs}
    \begin{tabular*}{\textwidth}{c@{\extracolsep{\fill}}c@{\extracolsep{\fill}}c@{\extracolsep{\fill}}c@{\extracolsep{\fill}}c@{\extracolsep{\fill}}c@{\extracolsep{\fill}}c@{\extracolsep{\fill}}c@{\extracolsep{\fill}}c@{\extracolsep{\fill}}c@{\extracolsep{\fill}}c@{\extracolsep{\fill}}}
		\hline 
		\hline \\[-6pt]
	    \multicolumn{11}{c}{(a) $\mu(\mathbf{X}) = \left( \V{\omega}^{\prime} \M{X} \right)^3$} \\[3pt]
	    \hline
		& & \multicolumn{4}{c}{TR} & \multicolumn{4}{c}{SS} & \\ 
		\cline{3-6} \cline{7-10} \\ [3pt] 
		Model & $\bm{\beta}$ & Bias & SD & SE & CP & Bias & SD & SE & CP & RE \\ 
		\hline
	    &0 &$-0.010$ &0.220 & 0.209 &0.94 &$-0.005$ &0.123 & 0.122 & 0.95 & 3.21 \\
		Linear &1 &$-0.021$ &0.329 & 0.347 &0.98 &$-0.008$ &0.182 & 0.173 & 0.93 & 3.28 \\
		&1 &$-0.020$ &0.355 &0.347 &0.97 &$-0.004$ &0.198 & 0.175 & 0.92 & 3.22 \\ [0.5ex] 
		\hline
		&0 &$-0.008$ &0.206 & 0.205 &0.94 &$\ \ 0.001$ &0.123 & 0.132 & 0.95 & 2.80
		\\
		Cubic &0.41 &$\ \ 0.002$ &0.352 &0.338 &0.95 &$-0.002$ & 0.212 & 0.193 & 0.95 & 2.74 \\
	    &0.81 &$\ \ 0.002$ &0.423 &0.390 &0.94 &$-0.010$ &0.239 & 0.212 & 0.91 & 3.14 \\ [0.5ex]
		\hline
		&0 &$-0.006$ &0.171 & 0.168 &0.96 &$\ \ 0.004$ &0.118 & 0.116 & 0.95 & 2.13 \\
		Sine &0.37 &$-0.007$ &0.282 &0.270 &0.94 &$-0.011$ & 0.170 & 0.158 & 0.91 & 2.74 \\
	    &0.37 &$\ \ 0.011$ &0.296 &0.272 &0.94 &$\ \ 0.011$ &0.176 & 0.161 & 0.92 & 2.82 \\ [0.5ex]
	    \hline
	    \hline \\
	    \multicolumn{11}{c}{(b) $\mu(\mathbf{X}) = \left( \V{\alpha}^{\prime} \M{X} \right) \left( 1 + \V{\omega}^{\prime}\M{X} \right)$} \\[3pt]
	    \hline
		& & \multicolumn{4}{c}{TR} & \multicolumn{4}{c}{SS} & \\
		\cline{3-6} \cline{7-10} \\ [3pt] 
		Model & $\bm{\beta}$ & Bias & SD & SE & CP & Bias & SD & SE & CP & RE \\ 
		\hline
	    &0 &$-0.017$ &0.242 & 0.230 &0.94 &$-0.007$ &0.114 & 0.116 & 0.95 & 4.53 \\
		Linear &1 &$-0.012$ &0.348 & 0.326 &0.94 &$\ \ 0.011$ &0.150 & 0.151 & 0.93 & 5.37 \\
		&1 &$-0.021$ &0.352 &0.347 &0.94 &$-0.011$ &0.163 & 0.154 & 0.91 & 4.66 \\ [0.5ex] 
		\hline
		&0 & $-0.005$ &0.236 & 0.223 &0.93 &$-0.026$ & 0.121 & 0.122 & 0.95 & 3.77 \\
		Cubic &0.41 &$-0.014$ &0.341 &0.313 &0.93 &$\ \ 0.003$ & 0.155 & 0.154 & 0.92 & 4.86 \\
	    &0.81 &$-0.004$ &0.432 &0.392 &0.91 &$-0.008$ &0.193 & 0.186 & 0.92 & 4.99 \\
		\hline
		&0 &$\ \ 0.003$ &0.210 & 0.202 &0.94 &$\ \ 0.009$ &0.117 & 0.113 & 0.95 & 3.14 \\
		Sine &0.37 &$\ \ 0.002$ &0.296 &0.289 &0.94 &$-0.004$ & 0.146 & 0.146 & 0.94 & 4.12 \\
	    &0.37 &$-$0.005 &0.300 &0.290 &0.93 &$\ \ 0.001$ &0.136 & 0.140 & 0.95 & 4.90 \\
	    \hline
\end{tabular*}
\vspace*{12pt}
\end{table}

\begin{table}
	\centering 
	\caption{Estimated parameters and their associated standard errors (SE) and p-values.}
	\label{table:RealData}
    \begin{tabular*}{\textwidth}{c@{\extracolsep{\fill}}c@{\extracolsep{\fill}}c@{\extracolsep{\fill}}c@{\extracolsep{\fill}}c@{\extracolsep{\fill}}c@{\extracolsep{\fill}}c@{\extracolsep{\fill}} c@{\extracolsep{\fill}} }
		\hline
		\hline
		& \multicolumn{3}{c}{TR} & \multicolumn{3}{c}{SS} \\
		\cline{2-4} \cline{5-7} \\ [1pt] 
	     Predictors & $\V{\beta}_{TR}$ & SE & p-value & $\V{\beta}_{SS}$ & SE & p-value \\
		\hline
	    Intercept & $\ \ 0.020$ &  0.136 & 0.883 &$-0.050$ &0.126 & 0.695 \\
		Mean MAP & $\ \ 0.043$ &0.094 & 0.640 & $\ \ 0.037$ &0.066  & 0.579 \\
		Mean heart rate  & $-0.096$ & 0.057 &  0.089 &$ - 0.104$ &0.050 & 0.040 \\
		[0.5ex] 
		\hline\\
	\end{tabular*}
\vspace*{12pt}	

\end{table}

\begin{table}
    \centering
    \caption{Treatment allocation given by SS and TR.} \label{table:allocation}
        \begin{tabular}{c@{\extracolsep{\fill}}c@{\extracolsep{\fill}}c@{\extracolsep{\fill}}c@{\extracolsep{\fill}}}
        \hline
                &             & \multicolumn{2}{c}{SS} \\
                & Treatment &  IV Fluid \ \ & Vasopressors \\[0.2ex]\hline
        \multirow{2}{*}{TR} \ \ & IV Fluid & 1430 & 29 \\
                & Vasopressors     & 459 &   1454 \\
        \hline
        \end{tabular} 
        \vspace{12pt}
\end{table}


\end{document}